\documentclass[runningheads,orivec]{llncs}
\usepackage[T1]{fontenc}
\usepackage{graphicx}
\usepackage{algorithm}
\usepackage[noend]{algpseudocode}
\usepackage{algorithmicx}
\usepackage{amsmath}
\usepackage{subcaption}
\usepackage{hyperref}
\usepackage[numbers,sort&compress]{natbib}
\begin{document}
\title{Federated Intrusion Detection System Based on Unsupervised Machine Learning}
\titlerunning{Federated IDS Based on Unsupervised ML}
%
\author{Maxime Gourceyraud \and Rim Ben Salem \and Christopher Neal \and Frédéric Cuppens \and Nora Boulahia Cuppens}
\authorrunning{M. Gourceyraud et al.}
%
\institute{Polytechnique Montréal, Montréal, QC, H3T 1J4, Canada \\
\email{\{maxime.gourceyraud, rim.ben-salem, christopher.neal, frederic.cuppens, nora.boulahia-cuppens\}@polymtl.ca}}
%
\maketitle              
\begin{abstract}
Recent Intrusion Detection System (IDS) research has increasingly moved towards the adoption of machine learning methods. However, most of these systems rely on supervised learning approaches, necessitating a fully labeled training set. In the realm of network intrusion detection, the requirement for extensive labeling can become impractically burdensome. Moreover, while IDS training could benefit from inter-company knowledge sharing, the sensitive nature of cybersecurity data often precludes such cooperation. To address these challenges, we propose an IDS architecture that utilizes unsupervised learning to reduce the need for labeling. We further facilitate collaborative learning through the implementation of a federated learning framework. To enhance privacy beyond what current federated clustering models offer, we introduce an innovative federated K-means++ initialization technique. Our findings indicate that transitioning from a centralized to a federated setup does not significantly diminish performance. 

\end{abstract}
\section{Introduction}
Cybercrime has become a major concern worldwide, posing significant economic problems. In fact, the World Trade Organization
(WTO) estimates that the cost of cybercrime reached \$8,150 billion USD in 2023 and could escalate to \$13,820 billion USD by 2028 \cite{OMC2024}. For organizations globally, detecting cyberattacks is crucial to mitigating their impact. Furthermore, with the increasing frequency of attacks, automating the detection process through  Intrusion Detection Systems (IDS) software has become essential. Traditionally, intrusion detection involves the use of signatures, where the IDS maintains a database of known attacks and compares incoming data against this database. A match with any known attack results in the data being labeled accordingly. While effective at identifying known threats, this method relies on static rules and is inadequate for adapting to new, evolving attacks. However, with advances in Machine Learning (ML), IDS systems have been increasingly been able to identify previously unseen attacks.

In recent years, ML-based IDS platforms have predominantly utilized supervised algorithms, to learn attack patterns by analyzing labeled data. However, the sheer volume of data generated by an organization's normal operations is too vast to label entirely, making it labor-intensive and resource-intensive. To circumvent this issue, unsupervised ML algorithms, which do not require extensive data labeling, can be employed. For instance, data clustering algorithms group data into partitions, or clusters, without prior knowledge of labels. A popular algorithm for data clustering is the K-Means algorithm, introduced by S. Lloyd \cite{Lloyd1982}, which creates clusters based on the proximity of data points to cluster centroids.

ML thrives on large, diverse datasets, which can pose challenges for organizations not primarily focused on data tasks. Collaboratively developing an IDS using varied data can leverage shared insights. However, traditional ML frameworks are centralized, requiring data to be consolidated at a single location. This centralization poses privacy risks and creates a single point of failure. For example, a compromised server holding data from all participants exposes extensive sensitive information. Conversely, decentralized data storage across departments or even different organizations force an attacker to breach each one individually, significantly reducing overall risk. The field of Federated Learning (FL), introduced by McMahan et al. \cite{McMahan2017}, addresses these issues by proposing a learning protocol that includes model learning at the client level and model aggregation at the server side, thus keeping the data localized.

This paper combines unsupervised and federated learning to develop an IDS. We validate this approach with experiments using the UNSW-NB15 \cite{Moustafa2015-nl} and CIC-IDS2017 \cite{Sharafaldin2018} datasets. We summarize the contributions of this work as follows:
\begin{itemize}
    \item We propose a federated learning-compatible architecture that integrates clustering algorithms for intrusion detection.
    \item We introduce a federated k-means++ initialization that is mathematically equivalent to its centralized counterpart.
    \item We present a method for computing a simplified silhouette score for clustering model selection within a federated framework.
\end{itemize}

The remainder of the paper is outlined as follows. A review of related work is provided in Section~\ref{sec:related_work}. The methodology of this work is detailed in Section~\ref{sec:methodology}. The experimental results of our approach are reported in Section~\ref{sec:results}. We discuss our findings in Section~\ref{sec:discussion} and conclude this work in Section~\ref{sec:conclusion}.

\section{Related Work}\label{sec:related_work}

This section presents an overview of clustering algorithms that are applicable in FL environments, followed by an introduction to relevant centralized and federated IDS frameworks.

\subsection{Clustering Algorithms}

\subsubsection{Client Clustering.}
One part of the federated clustering is not dedicated to cluster the clients' data but the clients themselves. This way, the clients are grouped in the aggregation step to enhance the performances of a supervised model. For example, CADIS \cite{Nguyen2023} uses the second last layer of locally learned deep learning models to calculate a customer similarity matrix. Using a layer of the model learned at the customer's site makes it possible to observe the distribution of data without having access to the data itself. When the models are aggregated by the server, the modifications made locally are weighted by the size of the  client's dataset and the inverse of the number of clients that are similar to it. This way, the weight of clients is reduced for those belonging to a large cluster and increased for those belonging to a small cluster.

\subsubsection{Federated Clustering.}
The type of federated clustering algorithms that we are interested in our study are those that cluster the clients' data. One of the first introduced was K-FED \cite{Dennis2021} and it is based on K-Means. The clients perform the clustering algorithm of Aswathi et al. \cite{Aswathi2012}. The server aggregates the centroids by using Lloyd's algorithm \cite{Lloyd1982}. K-FED works in only one round of communication. Each client must not have more than $\sqrt{k}$ of the overall $k$ clusters.

Two federated clustering algorithms are presented approximately at the same time. First, the Federated K-Means by Garst and Reinders \cite{Garst2023} uses a K-Means++ initialization \cite{Hruschka2004} at the client level. Then, the server aggregates the centroids by performing Lloyd's algorithm with each centroid weighted by the size of the cluster it represents. The clients improve the model by doing one step of Lloyd's algorithm on their data with the global centroids. A new round of communication begins after the server once again aggregates the clients' centroids. The algorithm stops after a fixed number of rounds of communication.

The other algorithm presented around the same period of time as \cite{Garst2023} is proposed by Holzer et al. \cite{Holzer2023}. This algorithm does not require that every client participates in the protocol at the same time. Each selected client updates the global centroids with one step of Lloyd's algorithm. The server aggregates the new centroids with a weighted average cluster by cluster. The weight of each centroid is the size of its cluster on the client side. This approach is similar to the one that is used in Tensorflow Federated \cite{BatchTff}. But Holzer et al. use a learning rate to bound the updates and use inertia to limit the fluctuations of the global model.

\subsection{Unsupervised IDS}
In this section, we highlight related IDS work based on unsupervised learning that utilizes either a centralized or federated learning approach, then provide the summary of their results in Table~\ref{tab:recap_ids_litterature}. We focus on works that utilize the same datasets as in this paper (i.e. UNSW-NB15 and CIC-IDS2017), as we feel this offers a fair comparison.

\subsubsection{Centralized IDS.}

Prasad et al. \cite{Prasad2020} propose a clustering algorithm that creates arbitrary shapes, not only convex shapes. To achieve this, their method initializes a large number of initial clusters. These clusters are grouped by similarity. They accompany their method with a selection of variables. They achieve respective accuracies of 0.8860 and 0.8828 on the UNSW-NB15 and CIC-IDS2017 datasets. 

For their anomaly detection, Yang et al. \cite{Yang2022} train an auto-encoder on normal data. In addition to detecting anomalous data with a threshold, the proposed method quantifies the normal distribution (i.e. on normal data) of hidden layers with their mean and variance. During the detection phase, the value of each of the hidden layers is compared with normality using the Mahalanobis distance \cite{Mahalanobis1936}. The final anomaly score is a linear combination of the reconstruction error and all the Mahalanobis distances of the hidden layers. Then, they determine a threshold for this score to differentiate the attacks from the normal traffic. With their method, the authors obtained an accuracy of $0.85 \pm 0.01$ and an $F_1$ score of $0.84 \pm 0.01$.

\subsubsection{Federated IDS.}

Md Tayeen et al. \cite{Tayeen2023} propose a federated learning algorithm using FedAvg \cite{McMahan2017}. They train an auto-encoder in a federated manner on normal data and then perform anomaly detection. However, the only parameters sent to clients are the weights of the layer preceding the latent space and those of the layer following it. Strictly speaking, federated training only concerns the latent space. The $F_1$ score obtained on UNSW-NB15 is $0.91 \pm 0.003$.

Grammenos et al. \cite{Grammenos2019FederatedPW} introduce a federated streaming and differentially private algorithm for computing
Principal Component Analysis (PCA). Federated-PCA estimates the rank of PCA with unknown distribution in a memory-limited setting. While the reported numerical simulations highlight the algorithm's potential, its performance when faced with outliers or missing data remains to be evaluated. The work of Nguyen et al. \cite{Nguyen_2024} is in the same vein as \cite{Grammenos2019FederatedPW} and uses Federated PCA on Grassmann Manifold for Internet of Things (IoT) anomaly detection. Their unsupervised framework for efficient host-based IDS is formulated as a consensus optimization problem with the aim of achieving privacy-preservation and communication-efficiency.

Aouedi et al. \cite{Aouedi2022} propose a supervised IDS enhanced with unsupervised learning. They train an auto-encoder in a federated way. Afterwards, they train on the server side a neural network with little labeled data. This is not exactly an unsupervised model but we find this work as important to highlight as the intrusion detection literature in this domain is sparse.

\begin{table}[H]
    \centering
    \begin{tabular}{ |c|c|c|c|c|}
    \hline
     \textbf{Reference} & \textbf{Dataset} & \textbf{Setup} & \textbf{Accuracy} & \textbf{$\mathbf{F_1}$} \\
     \hline
     Prasaad et al. \cite{Prasad2020} & CIC-IDS2017 & Centralized & $0.8860$ & $0.8828$ \\
     \hline
     Sirisha et al. \cite{Sirisha2021} (k-means) & CIC-IDS2017 & Centralized & $0.79$ & $0.88$ \\
     \hline
     Sirisha et al. \cite{Sirisha2021} (IForests) & CIC-IDS2017 & Centralized & $0.37$ & $0.91$ \\
     \hline
     Yang et al. \cite{Yang2022} & UNSW-NB15 & Centralized & $0.85$ & $0.84$ \\
     \hline
     Md Tayeen et al. \cite{Tayeen2023} & UNSW-NB15 & Federated & / & $0.91$ \\
     \hline
     Aouedi et al. \cite{Aouedi2022} & UNSW-NB15 & Federated & $0.8432$ & / \\
     \hline
\end{tabular}
    \caption{Summary of the performances of relevant work}
    \label{tab:recap_ids_litterature}
\end{table}

\section{Methodology}\label{sec:methodology}


\subsection{IDS architecture}

\begin{figure}
    \centering
    \includegraphics[width=12cm]{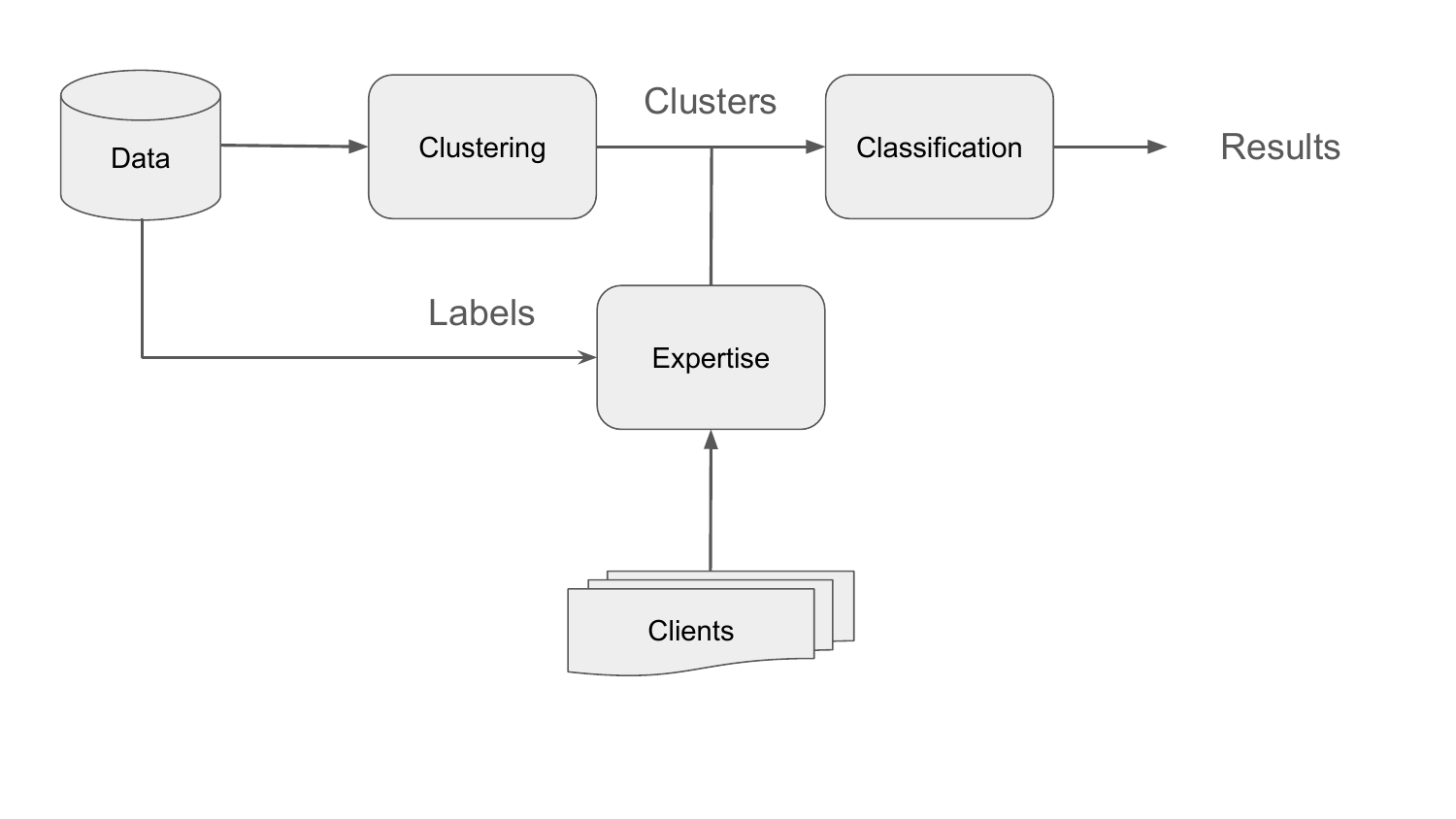}
    \caption{Representation of the proposed IDS}
    \label{fig:schema_ids}
\end{figure}

The proposed IDS architecture is represented in Figure \ref{fig:schema_ids}. It is composed of three main components: clustering, expertise, and classification. Data used by the IDS is first regrouped in clusters. Then, these clusters are analysed by the clients of the federated setup. Using the aforementioned expertise, each cluster is given a label according to its composition. Data is then classified as the label of the cluster it belongs to. The rest of this section describes each component.

\subsection{Federated Clustering} \label{sec:fed_kmeanspp}

The clustering part of the IDS, as shown in Figure \ref{fig:schema_ids} is either centralized or federated. To perform the clustering in a centralized manner, we use K-Means clustering \cite{Lloyd1982}. In the federated setup we used Federated K-Means clustering as proposed by Garst and Reinders \cite{Garst2023}. The behavior of the IDS in a centralized setup is used as a baseline for the performance of the federated setup.

However, the initialization of the Federated K-Means clustering of Garst and Reinders \cite{Garst2023} requires that every client participating in the protocol sends to the server the results of the k-means++ initialization \cite{Arthur2007-sf} on their data. Let $k \in \bbbn^*$ be the number of clusters required for the clustering and $N \in \bbbn^*$ be the number of clients. The protocol introduced by Garst and Reinders requires that the clients disclose a total of $kN$ entries. In this section, we propose the federated k-means++ initialization to enhance the privacy of the clustering algorithm.

\subsubsection{K-Means Properties.}

Before introducing the Federated K-means++ initialization, we need to reformulate the probability distributions used in K-Means++ \cite{Arthur2007-sf}. Let $(X_j)_{j \in \{1, ...,N\}}$ be the datasets of the clients $(\mathcal{C}_j)_{j \in \{1, ...,N\}}$ and $X = \bigcup\limits_{j  = 1}^N X_j$ the total dataset. Suppose that each $x \in X$ is either unique or there is a way to distinguish all the entries with the same values. For each $x \in X$, we have:

\begin{equation}
    P_{\mathcal{U}}(x) = \frac{1}{|X|} = \frac{|X_j|}{|X|} \times \frac{1}{|X_j|} = P_{\mathcal{U}}(C_j) P_{\mathcal{U}}(x|C_j)
  \label{eq:decentralized_uniform}
\end{equation}

Equation \ref{eq:decentralized_uniform} shows that the uniform distribution on a dataset $X$ can be equivalently expressed as a choice of a client $C_j$ that then realizes the uniform sampling on its own data. Similarly, the $D^2$ distribution introduced for the k-means++ initialization \cite{Arthur2007-sf} can be expressed as choice of a client and this client samples its data with a local $D^2$ distribution. This is shown by equation \ref{eq:decentralized_d2}. Let $c$ be the set of centroids at any moment. Let for all $j, Z_j = \sum_{x \in X_j} d(x,c)$ with $d(x,c) =  \min\limits_{i \in \{1, ...,|c|\}}(||x - c_i||^2_2)$ and $Z = \sum_{j = 1}^N Z_j = \sum_{x \in X} d(x,c)$. The $D^2$ probability distribution can be written as: 

\begin{equation}
    P_{\mathcal{D}^2}(x) = \frac{d(x, c)}{Z} = P(C_j) P_{\mathcal{D}^2}(x|C_j) = \frac{Z_j}{Z} \times \frac{d(x, c)}{Z_j}
  \label{eq:decentralized_d2}
\end{equation}

So the probability distributions used in the k-means++ initialisation can be written in a decentralized way for any distribution of data between the clients. Also the formulation of the probability distribtions does not allow the server to access the clients' data.

\subsubsection{Federated K-Means++ Initialization.} Algorithm \ref{alg:fed_kmeanspp} uses the properties discussed above to create a federated version of the K-Means++ algorithm equivalent to its centralized counterpart. With this algorithm, the server only accesses the centroids selected by the clients, and does not access to the rest of their data.

\begin{algorithm}
\caption{Federated K-Means++ Initialization}\label{alg:fed_kmeanspp}
\begin{algorithmic}[1]
\Require $k$, $\mathcal{C} = (\mathcal{C}_j)_j$, $(X_j)_j$

\State The clients send the size of their datasets to the server.
\State The server samples a client $\mathcal{C}_j$ in $\mathcal{C}$ with $P(\mathcal{C}_j) = \frac{|X_j|}{|X|}$
\State The client $\mathcal{C}_j$ samples the first centroid $c_1$ uniformly on $X_j$
\State $c \gets \{c_1\}$

\For{$i \in \{2,...,k\}$}
    \State The server samples a client $\mathcal{C}_{i'}$ in $\mathcal{C}$ with $\forall j, P(C_j) = \frac{Z_j}{Z}$
    \State The client $\mathcal{C}_{i'}$ samples $c_i$ where $P(c_i | \mathcal{C}_{i'}) = \frac{d(c_i, c)}{Z_{i'}}$
    \State $c \gets c \cup \{c_i\}$
    \EndFor

\Return $c$

\end{algorithmic}
\end{algorithm}

With Algorithm \ref{alg:fed_kmeanspp}, the clients only disclose the sampled data to the rest of the participants. If $k$ points are required, then the clients only reveal $k$ points and this initialization is equivalent to the centralized one as shown previously.

Algorithm \ref{alg:garst_kmeans++} explains how we insert the federated k-means++ initialization into Garst and Reinders' federated K-Means approach. $c^j$ are the centroids of the j-th client and $C^j$ are the clusters associated with them. We note $s^j$ the size of the clusters of the $j$-th client and $c^g$ the global centroids.

\begin{algorithm}
\caption{Federated K-Means with Federated K-Means++ initialization}\label{alg:garst_kmeans++}
\begin{algorithmic}
\Require $k$

\State \# Federated k-means++ initialization

\State $c_g \gets $ federated\_kmeans++($k, \mathcal{C}, (X_j)_j, $)

\For{round $r$}
    \State \textbf{For each client $\mathcal{C}_j \in \mathcal{C}$ :}
    \State\hspace{\algorithmicindent} $c^j \gets c^g$
    \State\hspace{\algorithmicindent} Determine $s^j$
    \State\hspace{\algorithmicindent} \# The client $\mathcal{C}_j$ ignores the centroids without data associated with.
    \State\hspace{\algorithmicindent} $s^j \gets C^j[s!=0 $ for $ s $ in $ s^j]$
    \State\hspace{\algorithmicindent} $k_j \gets $ size($s_j$)
    \State\hspace{\algorithmicindent} \# The clients compute one step of Lloyd's algorithm
    \State\hspace{\algorithmicindent} \# Lloyd's algorithm is applied with $c^j$ as intial centroids.
    \State\hspace{\algorithmicindent} $s^j, c^j \gets $ kmeans($X_j, k_j$, init = $c^j$)
    \State\hspace{\algorithmicindent} Send $s^j, c^j$ to the server.
    \State \textbf{At the server :}
    \State\hspace{\algorithmicindent} \# Centroids and size of clusters are concatenated.
    \State\hspace{\algorithmicindent} \# It creates a dataset on which the server applies k-means.
    \State\hspace{\algorithmicindent} $c \gets [c^1|c^2|...|c^{N}]$
    \State\hspace{\algorithmicindent} $s \gets [s^1|s^2|...|s^{N}]$
    \State\hspace{\algorithmicindent} \# Aggregation step. Lloyd's algorithm weighted by the clusters' size.
    \State\hspace{\algorithmicindent} $c^g \gets $ kmeans($c, k$, weights = $s$)
    \State\hspace{\algorithmicindent} Send $c^g$ to the clients.

\EndFor

\end{algorithmic}
\end{algorithm}

\subsection{Federated Simplified Silhouette}

 K-Means clustering has two main issues: the initialization and the choice of the number of clusters $k$. We cover the issue of initialization in Section \ref{sec:fed_kmeanspp}. Since clustering is used an unsupervised set of techniques, metrics such as accuracies are other performance metrics cannot be used. Now, we introduce the federated simplified silhouette score. The silhouette score \cite{Rousseeuw1987} is a metric to estimate how good a clustering is for each data point. To reduce the computation cost, Hruschka et al. \cite{Hruschka2004} proposed a simplified silhouette. We use the latter because it is less costly and because the classical silhouette score would require each client to know where each data point of the others are. This would undermine the efforts to improve privacy. For each $x \in X$, let's note $c(x)$ the centroid associated with $x$. Let $c$ be the set of the centroids and $d$ be a distance.  The simplified silhouette score is computed as follows:

 \[
\left \{
\begin{array}{c @{=} c}
    a(x) & d(x, c(x)) \\
    b(x) &  \min\limits_{i \in \{1,...,k\}, c_i \neq c(x)} d(x, c_i) \\
    s(x) & \frac{b(x) - a(x)}{max(a(x), b(x))}
\end{array}
\right. 
\]

We adapt the formula of $s$ to a federated setup. For each client $\mathcal{C}_j$, we have the average silhouette score of $C_j$ :

\begin{equation}
  \begin{split}
    \overline{S_{C_j}} &=  \frac{1}{|X_j|} \sum_{x \in X_j} s(x) \\
  \end{split}
  \label{eq:avg_silh_client}
\end{equation}

Then the server computes the average score weighted by the size of each dataset across the network : 

\begin{equation}
  \begin{split}
    S &=  \frac{1}{\sum_{j = 1}^{N} |X_j|} \sum_{j = 1}^N |X_j| \overline{S_{C_j}} \\
      &= \frac{1}{\sum_{j = 1}^{N} |X_j|} \sum_{j = 1}^N \sum_{x \in X_j} s(x) \\
    S &= \frac{1}{|X|} \sum_{x \in X} s(x)
  \end{split}
  \label{eq:fed_silh}
\end{equation}

Therefore, the average silhouette score computed is equivalent to if the server had direct access to all the data without actually accessing it. This is the average federated simplified silhouette score. The goal is to find the best value of $k$ by maximizing this score just like in a centralized setup. Throughout this paper, we use federated silhouette and federated simplified silhouette interchangeably.

\subsection{Binary Classification}
Once the clusters are formed, the IDS needs the clients to estimate the proportion of benign communications in each cluster. The server stores the proportion of benign communications in cluster $i$ of client $j$ in $p_{i,j}$. The clients also send to the server the size of the cluster in their dataset. Similarly, the size of cluster $i$ of the client $j$ is stored in $s_{i,j}$. The server then computes the proportion of benign communications across the network for each $i$ :

\begin{equation}
    P_i =  \frac{\sum_j p_{i,j} \times s_{i,j}}{\sum_j s_{i,j}}
  \label{eq:vote_bin}
\end{equation}

In a centralized setup, the data owner directly estimates $P$. Once $P$ is estimated, the rule we used is to consider the cluster $i$ benign if $P_i > 0.5$. Contrarywise, this cluster is considered as an attack cluster if $P_i \leq 0.5$. Once a cluster is classified as benign or attack, all the data associated with it are classified as such.

To measure the performances of the models, we set the positive class to be the attacks because it is what we want to detect. We use the classical metrics of accuracy, precision, recall and $F_1$ score.

\subsection{Datasets} \label{sec:datasets}
To evaluate the proposed architecture of the IDS and the federated algorithm proposed, we used two common datasets in intrusion detection: UNSW-NB15 and CIC-IDS2017. They are centralized datasets so we have to split them to create federated datasets, as explained in the following. 

\subsubsection{UNSW-NB15.}
The UNSW-NB15 dataset was built at the University of New South Wales to solve issues with previous datasets \cite{Moustafa2015-nl}. This dataset was designed to include realistic and various normal behavior and modern attacks. It is widely used in the field of intrusion detection to evaluate IDS. The dataset contains attacks that are relatively easy to detect such as DDoS, along with more complex attacks like exploits. 

To create a federated dataset from UNSW-NB15 we choose to regroup data with respect to the value of a certain variable. For this dataset, each client is composed by only a single class. For instance, one of the clients is entirely composed of the normal communications, one is entirely composed of DDoS attacks, etc. We recognize this does not represent an ideal setup for independent-identically distributed samples, however we see this as an acceptable compromise to create a federated dataset using UNSW-NB15.

\subsubsection{CIC-IDS2017.}
The CIC-IDS2017 \cite{Sharafaldin2018} dataset was created by the University of New Brunswick. The creators of this dataset insisted on simulating a realistic background traffic along with modern attacks. They use a diversity of protocols, devices, and operating systems to be as comprehensive and realistic as possible. Since the network was fully controlled by the researchers, each communication was labeled. The normal traffic is called "benign" and the attacks are labeled according to each type contained in the dataset. For instance, there are DDoS, Web attacks, etc.

For CIC-IDS2017, we create the federated dataset by regrouping client by destination IP of the communication. This way, each client is considered as a receiver of the communications and is closer to a realistic setup. 

\subsubsection{Data Preprocessing.}
Before data is used in our approach we perform the following preprocessing steps. We remove the dataset entries containing empty variables. We replace infinity values with minimum and maximum values of the corresponding variable. We replace the categorical variables by binary variables. Then, we normalize each variable so that they are limited to the $[0;1]$ interval. We apply the unsupervised variable selection based on value frequency \cite{Prasad2020}. Redundant data is removed from the dataset. Finally, we create a training set representing $80\%$ of the whole dataset and a test set representing $20\%$. The entire dataset is shuffled before creating the previously mentioned datasets. 

\subsubsection{Overview of Experimental Approach} \label{sec:xp_proto}
Here we outline the process used to train our federated IDS based on unsupervised learning: The algorithm is trained with values of the number of clusters $k$ ranging from $1$ to $300$ for CIC-IDS2017 and from $1$ to $70$ for UNSW-NB15. We vary the number of rounds of communications $r$ using the values $0$, $5$ and $10$. Then, the chosen value of $r$ is the minimum such that the silhouette curve has a maximum or local maximums. Following this, we choose the number of clusters such that it is the maximum, or a local maximum, of the simplified silhouette curve. The clients vote for the composition of each cluster, as shown previously. Based on this, the server aggregates the results. The class (attack/benign) of the cluster is determined according to the vote system chosen. Finally, each communication is labeled in accordance with the cluster it belongs to.

\section{Results}\label{sec:results}

\subsection{Model selection} \label{sec:silhouette_comp}

\begin{figure}
    \centering
     \begin{subfigure}[b]{0.45\textwidth}
         \includegraphics[width=\textwidth]{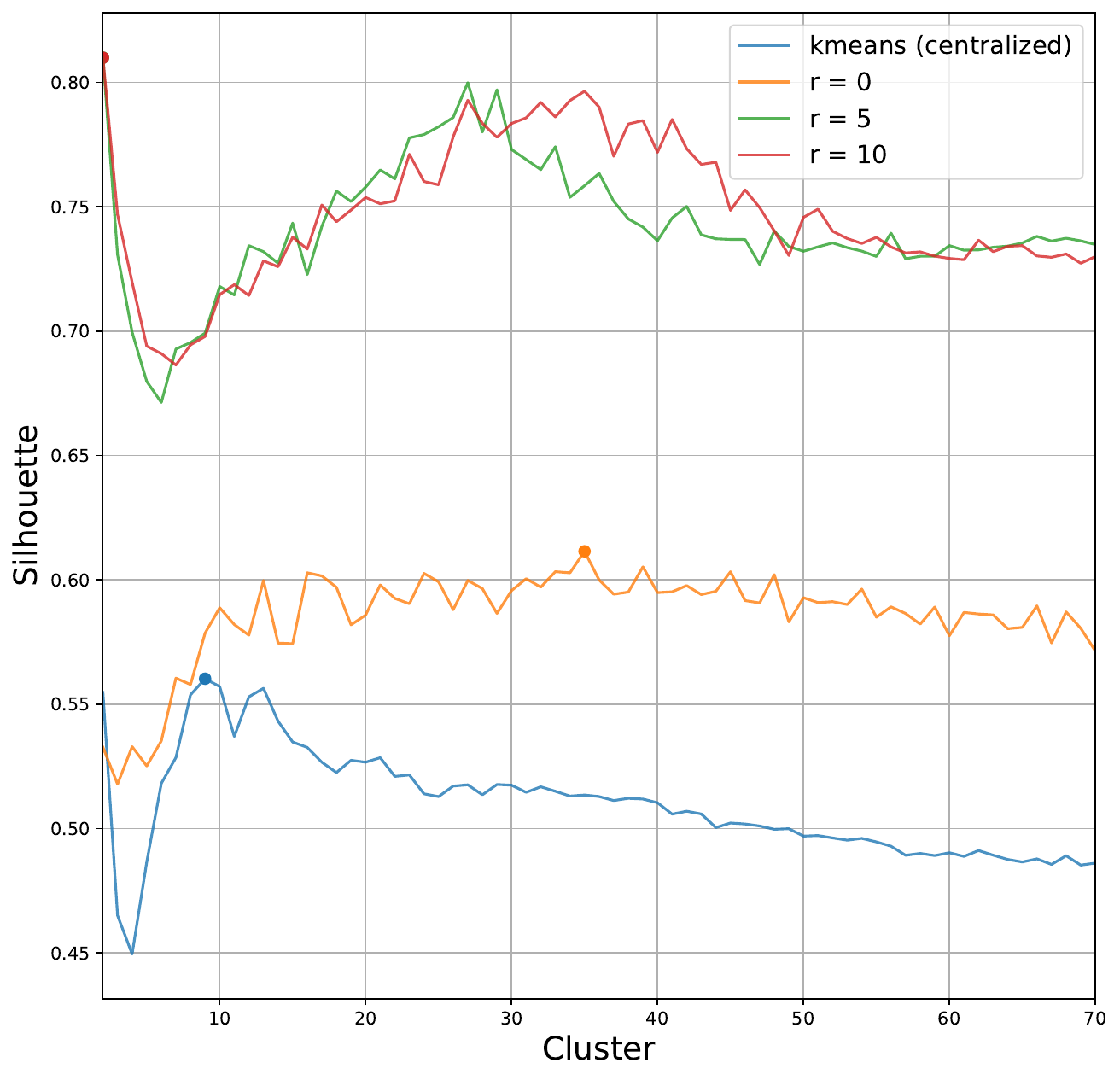}
         \caption{Evolution of the average silhouette score on UNSW-NB15 dataset}
         \label{fig:silh_comparison_unsw}
     \end{subfigure}
     \begin{subfigure}[b]{0.45\textwidth}
         \includegraphics[width=\textwidth]{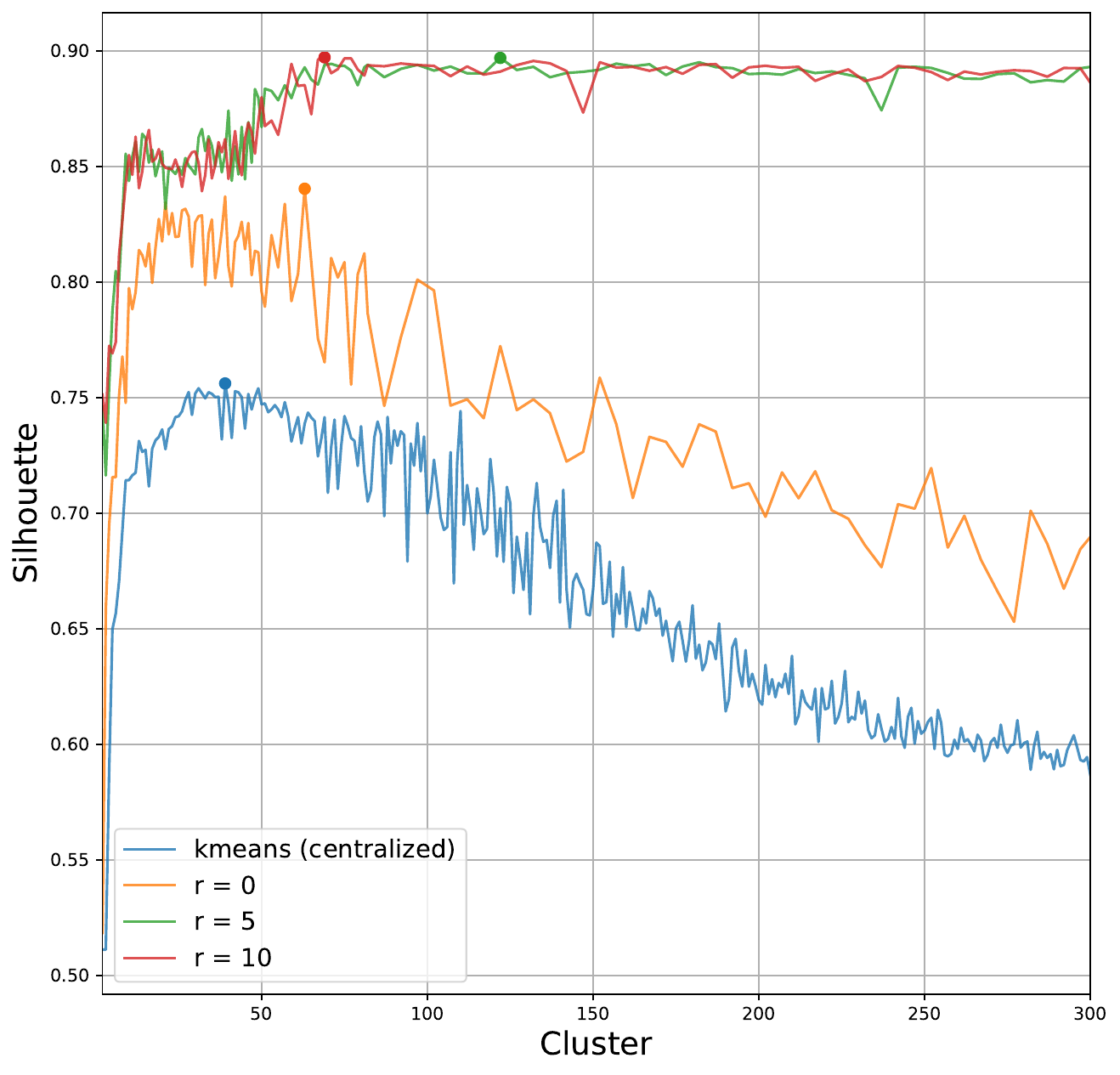}
         \caption{Evolution of the average silhouette score on CIC-IDS2017 dataset}
         \label{fig:silh_comparison_cic}
     \end{subfigure}

    \caption{Evolution of the average silhouette score as a function of the number of clusters (fed. K-Means with fed. K-Means++ init.).}
    \label{fig:silhouette_comparison}
\end{figure}

Figure \ref{fig:silhouette_comparison} presents the evolution of the silhouette score as a function of the number of clusters. Both UNSW-NB15 and CIC-IDS2017 are represented on these figures. The blue curves represent the silhouette score computed on the clusters generated with the centralized k-means algorithm. The other curves represent the evolution of the silhouette score as a function of $r$, the number of rounds of communication between the clients and the server.

\subsubsection{UNSW-NB15} For UNSW-NB15 (Figure \ref{fig:silh_comparison_unsw}), the centralized k-means curve and the curves for $r=5$ and $r=10$ share the same shape. Contrarywise, the curve $r=0$ seems flat. We tested the centralized version of k-means++ on the same data and it generated a curve similar to the one with $r=0$. So the "flat" shape is not an error of the federated k-means++. Even if the first three curves mentioned share the same shape, the maximums are not located at the same value of $k$, the number of clusters. In fact, the curves for $r=5$ and $r=10$ have a maximum around $k=30$ but the centralized k-means has a maximum at $k=9$. So the federated models used here shifts the maximum of the silhouette score to higher values of $k$.

The curve with $r=0$ shows no clear maximum. So we chose the number of round of communcations tested that produces a curve with a clear maximum. In that case, we chose $r=5$. We then select the maximum of the curve which is $k=27$.

\subsubsection{CIC-IDS2017} For CIC-IDS2017 (subfigure \ref{fig:silh_comparison_cic}), we can see that the silhouette curves for the centralized k-means and the federated k-means++ ($r=0$) share the same shape. Contrary to UNSW-NB15, the curves for $r=5$ and $r=10$ do not look like the centralized one.  These two last curves do not have a clear maximum.

The only federated curve whose shape resembles that of the centralized curve is with $r=0$. So we chose this number of round of communications and the maximum of this curve which is $k=63$.

\subsubsection{Summary of Model Selection.} 
We can see on figure \ref{fig:silhouette_comparison} that when $r$ increases, the silhouette score increases. It could show an improvement in the quality of clustering as the silhouette score measures this. However, the shapes of the curves in figure \ref{fig:silh_comparison_cic} for $r>0$ are clearly not the kind that is expected. Indeed, a silhouette curve is expected to have at least one local maximum and to decrease as $k$ increases. Here it is clearly not the case even for $r=0$ and in the centralized setup, the curves have a maximum before $k=300$. This shows that the federated iterations do not behave as expected. The suboptimality is also present on the UNSW-NB15 dataset, the maximums of the curves $r>0$ are shifted to higher values of $k$ compared to the centralized setup.

So we understand that the federated models studied generate suboptimal results with respect to the silhouette score. We know that the initialization we proposed is equivalent to its centralized counterpart. But we have no \textit{a priori} guarantee that the aggregation function from Garst and Reinders \cite{Garst2023} that we used is also equivalent to the step of Lloyd's algorithm. This can explain why the results are not as satisfactory as the centralized ones.

\subsection{Evolution of Performance Metrics} \label{sec:evolution_perfs}
\begin{figure}
    \centering
     \begin{subfigure}[b]{0.45\textwidth}
         \includegraphics[width=\textwidth]{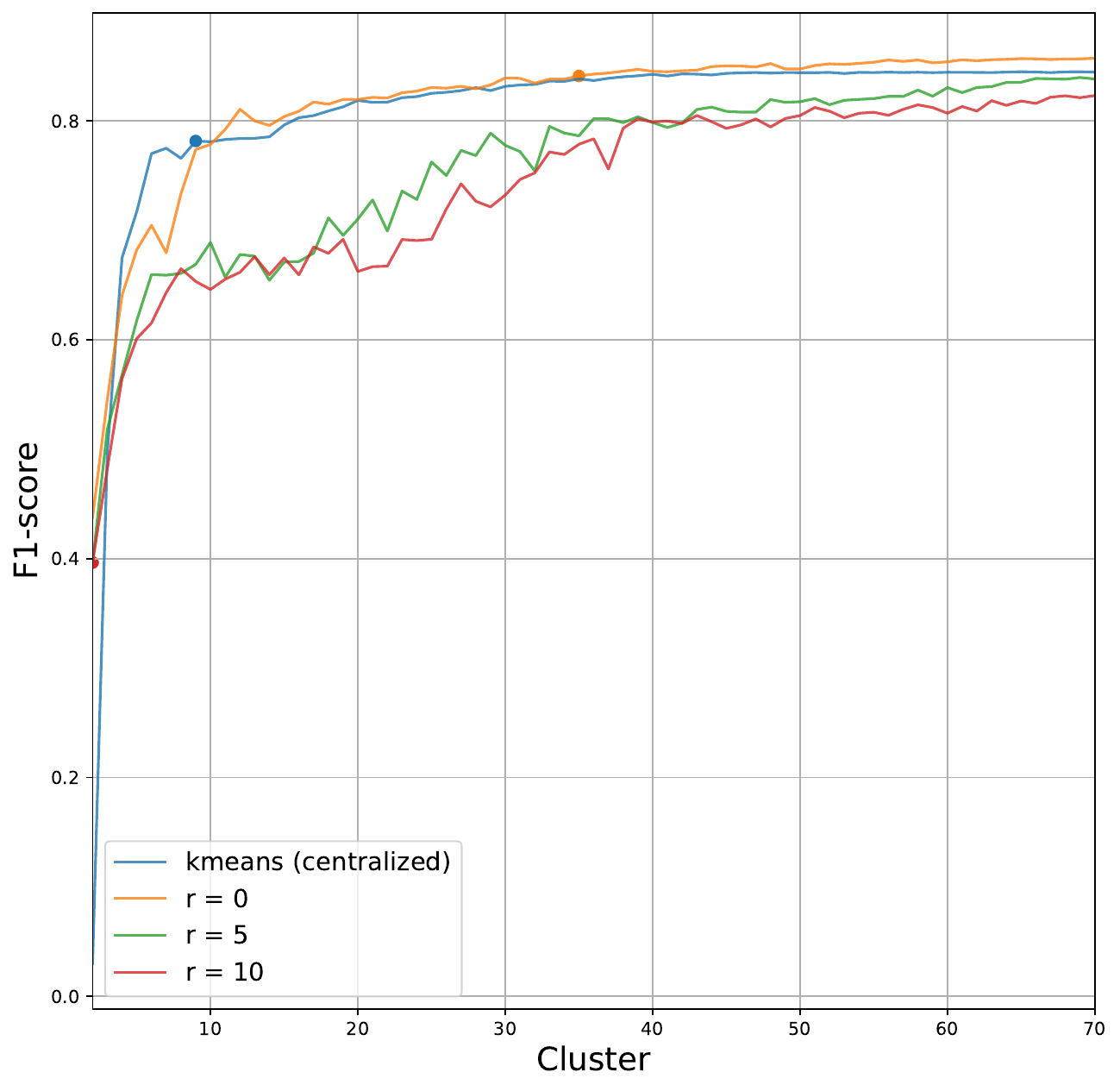}
         \caption{Evolution of $F_1$ score on UNSW-NB15 dataset}
         \label{fig:f1_comparison_unsw}
     \end{subfigure}
     \begin{subfigure}[b]{0.45\textwidth}
         \includegraphics[width=\textwidth]{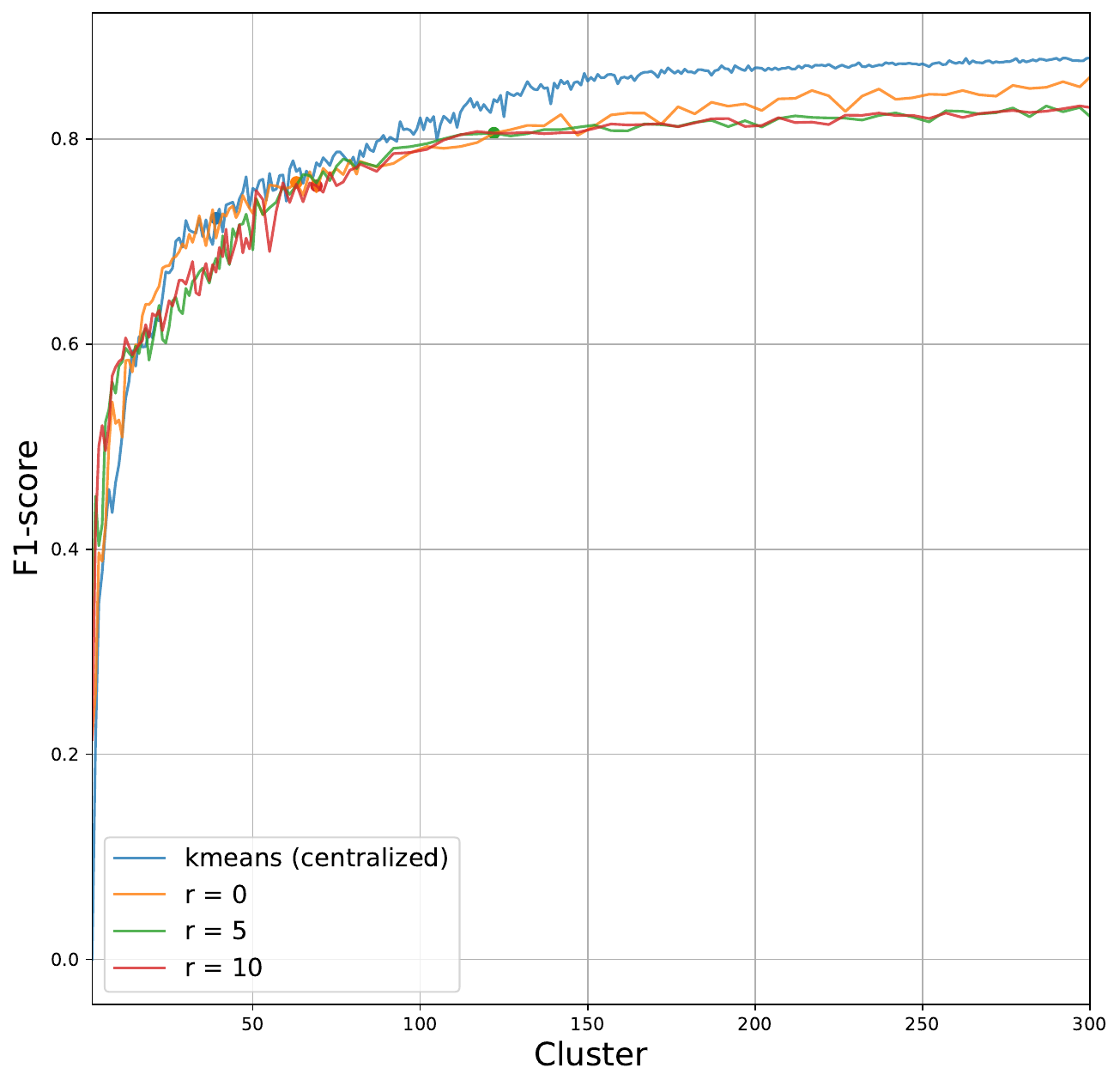}
         \caption{Evolution of $F_1$ score on CIC-IDS2017 dataset}
         \label{fig:f1_comparison_cic}
     \end{subfigure}

    \caption{Evolution of the average $F_1$ score as a function of the number of clusters (fed. K-Means with fed. K-Means++ init.).}
    \label{fig:f1_comparison}
\end{figure}

Figure \ref{fig:f1_comparison} presents the evolution of the $F_1$ score as a function of the number of clusters. Similar to figure \ref{fig:silhouette_comparison}, this figure shows the evolution of the $F_1$ score for different values of $r$ for the federated K-Means with federated K-Means++ initialization.

\subsubsection{UNSW-NB15} For UNSW-NB15 (figure \ref{fig:f1_comparison_unsw}), we can see that all curves are increasing in trend. There are two groups of curves. The first one is composed of the centralized algorithm and the federated one with $k=0$. This group shows better performances. The other one is composed of the curves with $r$ set to $5$ and $10$. Even though they get closer to the other two as $k$ increases, they are almost always below them.

\subsubsection{CIC-IDS2017} For CIC-IDS2017 (figure \ref{fig:f1_comparison_cic}), all curves are increasing in trend just like for UNSW-NB15. Contrary to the latter, the federated curves on the former are closer and the centralized one is clearly above the others starting from $k=50$. The curve $r=0$ is above the other two federated ones between $k=20$ and $k=70$ and then after $k=160$.

\subsubsection{Summary of Performance Evolution.} For both datasets, the performances of the centralized k-means are often better than the federated ones for fixed $k$. It means that the more the clusters, the better the performance. But more clusters also mean more expertise from the clients. If this expertise is provided by cybersecurity analysts, it can be tedious to label too many clusters. Moreover, we used unsupervised learning to limit the labeling. So the participants should be aware of such behavior before using the proposed IDS.

We also notice that the performances for $r=0$ are equivalent or better than the other two federated scenario. In terms of performances, it is better to keep $r$ as low as possible. So iterating on the centroids from the federated k-means++ initialization is not beneficial in term of performances. This is in line with the suboptimality of the aggregation function that we pointed in section \ref{sec:silhouette_comp}

\subsection{Performance Comparison} \label{sec:perf_comparison}

Table \ref{tab:comparaison_modeles_uns} and \ref{tab:comparaison_modeles_cic} compare the performances of the K-Means algorithm, federated K-Means \cite{Garst2023} and the federated K-Means with the proposed federated K-Means++ initialization. The results are computed using UNSW-NB15 and CIC-IDS2017 respectively.

\subsubsection{UNSW-NB15 dataset.} 
As we can see in Table \ref{tab:comparaison_modeles_uns}, the federated models have better accuracies and precisions than the centralized model. It means they produce less false positives. But both of them present a lower recall, which means that they produce more false negatives. The performances of the federated models should be compared with the number of clusters these methods need. Indeed, both of the federated algorithms require significantly more clusters than the centralized one.

\begin{table}[H]
    \centering
    \begin{tabular}{ |c|c|c|c|c|c|c|}
    \hline
     Model & Accuracy & Precision  & Recall  & $F_1$ & $k$ & $r$ \\
     \hline
     k-means (centralized) & $0.7659$ & $0.6640$ & $0.9500$ & $0.7817$ & $9$ & $/$ \\
     \hline
     Garst\&Reinders & $0.7706$ & $0.6904$ & $0.9201$ & $0.7889$ & $15$  & $0$ \\
     \hline
     Fed. k-means \& fed. k-means++ & $0.7743$ & $0.7121$ & $0.8456$ & $0.7731$ & $27$ & $5$ \\
     \hline
\end{tabular}
    \caption{Performance comparison between the models with selected hyperparameters (UNSW-NB15)}
    \label{tab:comparaison_modeles_uns}
\end{table}

\subsubsection{CIC-IDS2017 dataset} 
In Table \ref{tab:comparaison_modeles_cic}, we can see that the federated models have comparable or better accuracies, recalls and $F_1$ scores than the classical, centralized one. However, they have lower precisions and they need more clusters. We also notice that the selected $r$ is $0$ for both federated algorithms. As discussed in sections \ref{sec:silhouette_comp} and \ref{sec:evolution_perfs}, it seems that it is not always relevant to chose $r>0$. For our study, $r=0$ with the federated k-means of Garst and Reinders means that there is only one aggregation step after the local k-means++ initializations.

\begin{table}[H]
    \centering
    \begin{tabular}{ |c|c|c|c|c|c|c|}
    \hline
     Model & Accuracy & Precision  & Recall  & $F_1$ & $k$ & $r$ \\
     \hline
     k-means (centralized) & $0.8958$ & $ {0.7505}$ & $0.6969$ & $0.7227$ & $ {39}$ & $/$ \\
     \hline
     Garst\&Reinders & $0.9121$ & $0.7041$ & $ {0.9554}$ & $0.8107$ & $83$ & $0$ \\
     \hline
     Fed. k-means \& fed. k-means++ & $0.8947$ & $0.6765$ & $0.8613$ & $0.7578$ & $63$ & $0$ \\
     \hline
\end{tabular}
    \caption{Performance comparison between the models with selected hyperparameters (CIC-IDS2017)}
    \label{tab:comparaison_modeles_cic}
\end{table}

\section{Discussion}\label{sec:discussion}

In Section \ref{sec:silhouette_comp} and \ref{sec:evolution_perfs}, we notice that there is a suboptimality in the aggregation process. Indeed, the behavior of the model gets worse as the number of rounds increases. However, the performances of the studied models are better in terms of accuracy than the clustering algorithms from Table \ref{tab:recap_ids_litterature}.

Even though the rounds of communication do not increase the performances of the model, the performances obtained in a centralized setup or in a federated one are similar or better. The federated models require more clusters to achieve the same performances. So the federated clustering models studied are less efficient but the performances temper the suboptimality comment. It leaves room for future research.

Since the model detects attacks, it shows that the clients did collaborate during the training steps. For example in UNSW-NB15, one of the clients has only normal communications. But with the centroids learned across the network it is now capable of detecting attacks even though they were not previously seen. Similarly, with the federated dataset we created for CIC-IDS2017, $203$ datasets on $212$ are only composed of normal communications. The learning process is still capable of finding attacks. 

\section{Conclusion}\label{sec:conclusion}
In this paper we propose an IDS architecture that uses clustering algorithms, centralized or not. We improved the privacy of the clients' data compared to the federated k-means from Garst and Reinder with our federated k-means++ initialization without any loss of performance overall. Moreover, we found that the IDS and the learning process we described enable collaboration among the clients without sharing the datasets to the server. To select the federated clustering models, we proposed an average federated simplified silhouette score. We showed it is the same as the centralized one.

Future research should focus on refining aggregation techniques to enhance the performance of federated models, potentially leading to more robust and scalable cybersecurity solutions in decentralized networks. This direction not only promises improvements in IDS efficacy but also broadens the potential for cross-organizational collaboration without compromising sensitive information.



%
%
%
 \bibliographystyle{splncs04}
 \bibliography{bibliography}

\end{document}